\documentclass[12pt]{iopart}

\usepackage[T1]{fontenc}       
\usepackage{graphicx,color}
\usepackage{array}
\usepackage{textcomp}
\usepackage{multirow}
\usepackage{subscript}
\usepackage{inputenc}
\usepackage{dcolumn}

\usepackage{citesort}

\newcommand{\tabularnewline}{\\}
\newcommand{\mos}{MoS$_2$}
\newcommand{\Mos}{MoS\textsubscript{2} }
\newcommand{\wmk}{Wm$^{-1}$K$^{-1}$}

\begin{document}
	
\title[ ]{Strongly Tunable Anisotropic Thermal Transport in MoS\textsubscript{2} by Strain and Lithium Intercalation: First--Principles Calculations}

\author{Shunda Chen$^1$, Aditya Sood$^{2,3,6}$, Eric Pop$^{2,4,5}$, Kenneth E. Goodson$^3$, Davide Donadio$^{1}$}
\address{$^1$Department of Chemistry, University of California Davis, One Shields Ave. Davis, CA 95616, USA}
\address{$^2$Department of Electrical Engineering, Stanford University, Stanford, CA 94305, USA}
\address{$^3$Department of Mechanical Engineering, Stanford University, Stanford, CA 94305, USA}
\address{$^4$Department of Materials Science and Engineering, Stanford University, Stanford, CA 94305, USA}
\address{$^5$Precourt Institute for Energy, Stanford University, Stanford, CA 94305, USA}
\address{$^6$Present address: Stanford Institute for Materials and Energy Sciences, SLAC National Accelerator Laboratory, Menlo Park, CA 94025, USA}
\ead{\mailto{shdchen@ucdavis.edu}, \mailto{ddonadio@ucdavis.edu}}

\begin{abstract}
The possibility of tuning the vibrational properties and the thermal conductivity of layered van der Waals materials either chemically or mechanically paves the way to significant advances in nanoscale heat management. 
Using first-principles calculations we investigate the modulation of heat transport in \Mos by lithium intercalation and cross-plane strain.
We find that both the in-plane and cross-plane thermal conductivity ($\kappa_{{r}}$, $\kappa_{{z}}$) of \Mos are extremely sensitive to both strain and electrochemical intercalation.
Combining lithium intercalation and strain, the in-plane  and cross-plane thermal conductivity can be tuned over one and two orders of magnitude, respectively.
Furthermore, since $\kappa_{{r}}$ and $\kappa_{{z}}$ respond in different ways to intercalation and strain, the thermal conductivity anisotropy can be modulated by two orders of magnitude. 
The underlying mechanisms for such large tunability of the anisotropic thermal conductivity of \Mos are explored by computing and analyzing the dispersion relations, group velocities, relaxation times and mean free paths of phonons. 
Since both intercalation and strain can be applied reversibly, their stark effect on thermal conductivity can be exploited to design novel phononic devices, as well as for thermal management in \mos-based electronic and optoelectronic systems. 
\end{abstract}

\submitto{\TDM}
\maketitle

\section{Introduction}


Among two-dimensional {(2D)} layered materials, transition metal dichalcogenides (TMDCs) are of special interest for next-generation electronic and optoelectronic devices. In fact, in contrast to graphene, {many TMDCs} are semiconductors with considerable band gap \cite{chhowalla_chemistry_2013,wang_electronics_2012,butler_progress_2013,Lee_review_2017}.
Among TMDCs, molybdenum disulphide (\mos) has been explored most intensively \cite{splendiani_emerging_2010,kis_single_2011,butler_progress_2013,wang_electronics_2012,sundaram_electroluminescence_2013,lee_mos2_2012,yin_single-layer_2012,tsai_monolayer_2014,Lee_review_2017}.
Its mechanical flexibility makes it a compelling semiconducting material for flexible electronics \cite{pu_highly_2012,chang_high-performance_2013,cheng_few-layer_2014,wan_flexible_2015},
and its large interlayer separation provides ideal space to intercalate guest species, such as alkali metal ions. Lithium intercalation of MoS\textsubscript{2}, has been reported to enhance the optical transmission and increase electrical conductivity due to changes in the electronic band structure and the
injection of free carriers \cite{xiong_li_2015}. Reversible intercalation and deintercalation of Li in MoS\textsubscript{2} opens a route for novel applications in batteries and supercapacitors \cite{rasamani_interlayer-expanded_2017}. 
%

Thermal management in such applications, both at the nanoscale and at the macroscopic scale is a crucial issue \cite{pop_energy_2010}. 
As in other layered van der Waals materials, heat transport in \Mos is strongly anisotropic featuring high conductivity in plane and low conductivity across the layers.
In spite of its importance, the estimates of thermal conductivity of \Mos reported are in a wide range. 
The experimental estimates for in-plane thermal conductivity ($\kappa_{{r}}$) at room temperature vary from {13.3} to 110 \wmk\  \cite{sahoo_temperature-dependent_2013,yan_thermal_2014,jo_basal-plane_2014,liu_measurement_2014,zhang_measurement_2015,bae_in-plane-2017,jiang_time-domain_2017}, and those of cross-plane thermal conductivity ($\kappa_{{z}}$) vary from 2.0 to 5.3 \wmk\  \cite{liu_measurement_2014,muratore_thermal_2014,jiang_time-domain_2017}.
Such discrepancies may stem either from different  quality of samples 
or from different experimental conditions and measurement techniques. 

Even though it was suggested that intercalation may produce substantial reduction of thermal conductivity \cite{wan_flexible_2015,cho_electrochemically_2014}, fundamental aspects of thermal transport in intercalated \Mos are still not well understood.
Recent measurements indicate that lithiation reduces $\kappa_{{z}}$ of a \Mos thin-film  by a factor of two, while the effect on bulk samples is non-monotonic as a function of the concentration of Li \cite{zhu_tuning_2016}. At a fractional Li composition $x$ (in Li\textsubscript{x}MoS\textsubscript{2}) of 0.86, $\kappa_{{z}}$ of bulk \Mos is reduced by 1.25 times, and $\kappa_{{r}}$ is reduced by about 1.31 times \cite{zhu_tuning_2016}. In addition, thermal conductance across tens of nanometer thick \Mos films can be reversibly reduced by $\sim 10$ times through electrochemical lithiation \cite{ECT-Transistor}.

Atomistic simulations could provide a rationale for such experimental findings, and help resolving  discrepancies in experiments.
Heat transport in \Mos has been addressed theoretically by both first-principles anharmonic lattice dynamics and molecular dynamics (MD) simulations. 
First-principles calculations can be transferable and predictive, but the computation of anharmonic force constants and the solution of the linearized Boltzmann transport equation (BTE) need to  satisfy tight convergence criteria.
Former first-principles BTE calculations give $\kappa_{{r}}$ between 83 to 103 \wmk\  \cite{gandi_thermal_2016,gu_layer_2016,lindroth_thermal_2016}, and $\kappa_{{z}}$ from 2.3 to 5.1 \wmk\ \cite{gandi_thermal_2016,gu_layer_2016,lindroth_thermal_2016}.  {Non-monotonic changes of $\kappa$ as a function of partial lithium intercalation were recently reproduced by first-principles calculations on finite thickness open systems \cite{Luisier2018}.}
MD simulations, in turn, depend on the quality of the empirical potentials utilized, and provide too large a range of estimates for $\kappa_{{r}}$ from 1.35 to 110 \wmk\  \cite{varshney_md_2010,liu_phonon_2013,jiang_molecular_2013,wei_phonon_2014,ding_-plane_2015,kandemir_thermal_2016,hong_thermal_2016}, whereas for $\kappa_{{z}}$ values are between 4 and 6.6 \wmk\ \cite{varshney_md_2010,wei_cross-plane_2015}.
Interestingly, MD simulations predicted a large modulation of $\kappa_{{z}}$ as a function of cross-plane strain \cite{ding_-plane_2015}, but it is necessary to verify whether this result is independent of the adopted empirical potential, so to provide quantitative predictions that would aid the design of new \mos-based materials and devices.  
 
While the possibility of modulating the thermal conductivity of \Mos  is very appealing for applications in nanoscale electronic and thermal devices with tunable thermal functionality \cite{cahill_APR_2014,wang_physical_2015,chiritescu_ultralow_2007,mavrokefalos_-plane_2007,luckyanova_anisotropy_2013,muratore_thermal_2014,luo_anisotropic_2015,renteria_strongly_2015,chen_anisotropy_2016,chen_supersonic_2017,gao_lattice_2016,kang_ionic_2017}, a compelling understanding of the combined effects of strain and electrochemical intercalation is still lacking.

To fill this gap, we investigate the phonon properties and thermal conductivity of pristine and lithiated \Mos (LiMoS\textsubscript{2}), and we systematically probe the influence of cross-plane strain effects, by first-principles calculations.
We show that both the in-plane and cross-plane thermal conductivity of \Mos can be tuned substantially by lithium intercalation and cross-plane strain. 
For both pristine MoS\textsubscript{2} and LiMoS\textsubscript{2},
the cross-plane thermal conductivity is strongly enhanced by compressive strain and decreased by tensile strain. In \Mos such variation amounts to up to two orders of magnitude for the strain range investigated.
Furthermore we show that Li intercalation produces a more than seven-fold reduction of $\kappa_{{r}}$, while $\kappa_{{z}}$ is halved. 	
The different response of the in-plane and cross-plane components of $\kappa$ may be exploited to modulate the anisotropy ratio, so to achieve more or less directional heat dissipation. 



We compute the thermal conductivity of pristine hexagonal trigonal prismatic (2H)-MoS\textsubscript{2} and LiMoS\textsubscript{2} by solving the linearized phonon BTE, with harmonic and anharmonic force constants computed by first-principles density functional theory (DFT). DFT calculations are performed using both the local density approximation (LDA) and a van der Waals functional (vdW-DF) with consistent exchange (vdW-DF-cx) \cite{berland_van_2014}.
This first-principles BTE (FP-BTE) approach \cite{broido_intrinsic_2007} has proven accurate and predictive for a large variety of systems, including 2D and van der Waals layered materials \cite{ward_ab_2009,ward_intrinsic_2010,fugallo_ab_2013,garg_role_2011,lindsay_thermal_2012,lindsay_first-principles_2013,bonini_acoustic_2012,fugallo_thermal_2014,zeraati_highly_2016,lindroth_thermal_2016}.
As a result of linearizing and solving the BTE, the lattice thermal conductivity tensor $\kappa$ is expressed as:
\begin{equation}
\kappa_{\alpha\beta}=\frac{1}{V}\sum_{\lambda}\hbar\omega_{\lambda}\frac{\partial f}{\partial T}v_{\lambda}^{\alpha}v_{\lambda}^{\beta}\tau_{\lambda}^{\beta}\ ,\label{Eq.kappa}
\end{equation}
 where $f$ is the Bose-Einstein distribution function, $\omega_{\lambda}$ is the phonon angular frequency, $v_{\lambda}^{\alpha (\beta)}$ is the group velocity component along the $\alpha (\beta)$-direction, and 
$\tau_{\lambda}^{\beta}$ is the relaxation time for phonons with polarization $\lambda$ propagating in the direction $\beta$. Self-consistent BTE entails a directional dependence of phonon relaxation times, as the direction of the heat flux determines different shifts in the phonon population and therefore different scattering efficiency \cite{ward_ab_2009,zeraati_highly_2016}.
In the calculation of phonon relaxation times we consider intrinsic three-phonon scattering processes and extrinsic isotopic mass scattering \cite{tamura_isotope_1983}.  The natural isotopic distributions of Mo, S, and Li are considered \cite{berglund_isotopic_2011}  {and results are compared to those obtained for isotopically pure systems, so as to assess the effect of isotopic scattering.}  {Systems are considered in the infinite periodic bulk limit with no boundary scattering}.

Following Eq.~\ref{Eq.kappa}, FP-BTE not only provides a reliable estimate of $\kappa$, but it also allows one to resolve the contribution to $\kappa$ in terms of phonon frequency, polarization and mean free path (MFP), thus uncovering the mechanistic details of heat transport in crystalline materials. 
Our accurate, well converged, parameter-free first-principles calculations predict a strong modulation of thermal conductivity of layered MoS\textsubscript{2} by mechanical strain and lithium intercalation, and provide both a detailed microscopic interpretation of recent experiments and reliable benchmarks for future ones.

\section{Results and discussion}
\begin{table*}[htb]
\begin{tabular}{ccccccc}
\hline 
\hline
\multirow{2}{*}{Method} & \multicolumn{2}{c}{(2H)-MoS\textsubscript{2}} & \multicolumn{2}{c}{(1T)-LiMoS\textsubscript{2}} & \multirow{2}{*}{a-axis expansion} & \multirow{2}{*}{c-axis expansion}\tabularnewline
\cline{2-5} 
 & a  & c & a  & c &  & \tabularnewline
\hline 
LDA (This work) & 3.142 & 12.053 & 6.776 & 6.054 & 7.8\% & 0.46\%\tabularnewline
vdW-DF-cx (This work) & 3.158 & 12.284 & 6.789 & 6.229 & 7.5\% & 1.4\%\tabularnewline
LDA (Ref. \cite{gu_layer_2016}) & 3.14 & 12.05 & - & - & - & -\tabularnewline
PBE (DFT-D3) (Ref. \cite{gandi_thermal_2016}) & 3.157 & 12.225 & - & - & - & -\tabularnewline
vdW-DF-cx (Ref. \cite{lindroth_thermal_2016}) & 3.152 & 12.291 & - & - & - & -\tabularnewline
Exp. (Ref. \cite{petkov_structure_2002}) & 3.169 & 12.322 & - & - & - & -\tabularnewline
Exp. (Ref. \cite{mulhern_lithium_1989}) & - & - & 6.798 & 6.262 & 7.3\% & 1.6\%\tabularnewline 
Exp. (Ref. \cite{zhu_tuning_2016}) & - & 12.32 & - & 6.19 & - & 0.5\%\tabularnewline
\hline 
\hline
\end{tabular}
\caption{Equilibrium lattice parameters (in units of Angstrom) of pristine (2H)-MoS\textsubscript{2} and (1T)-LiMoS\textsubscript{2}.}
\label{table}
\end{table*}

\begin{figure*}[t]
\begin{center}
	\includegraphics[scale=0.28]{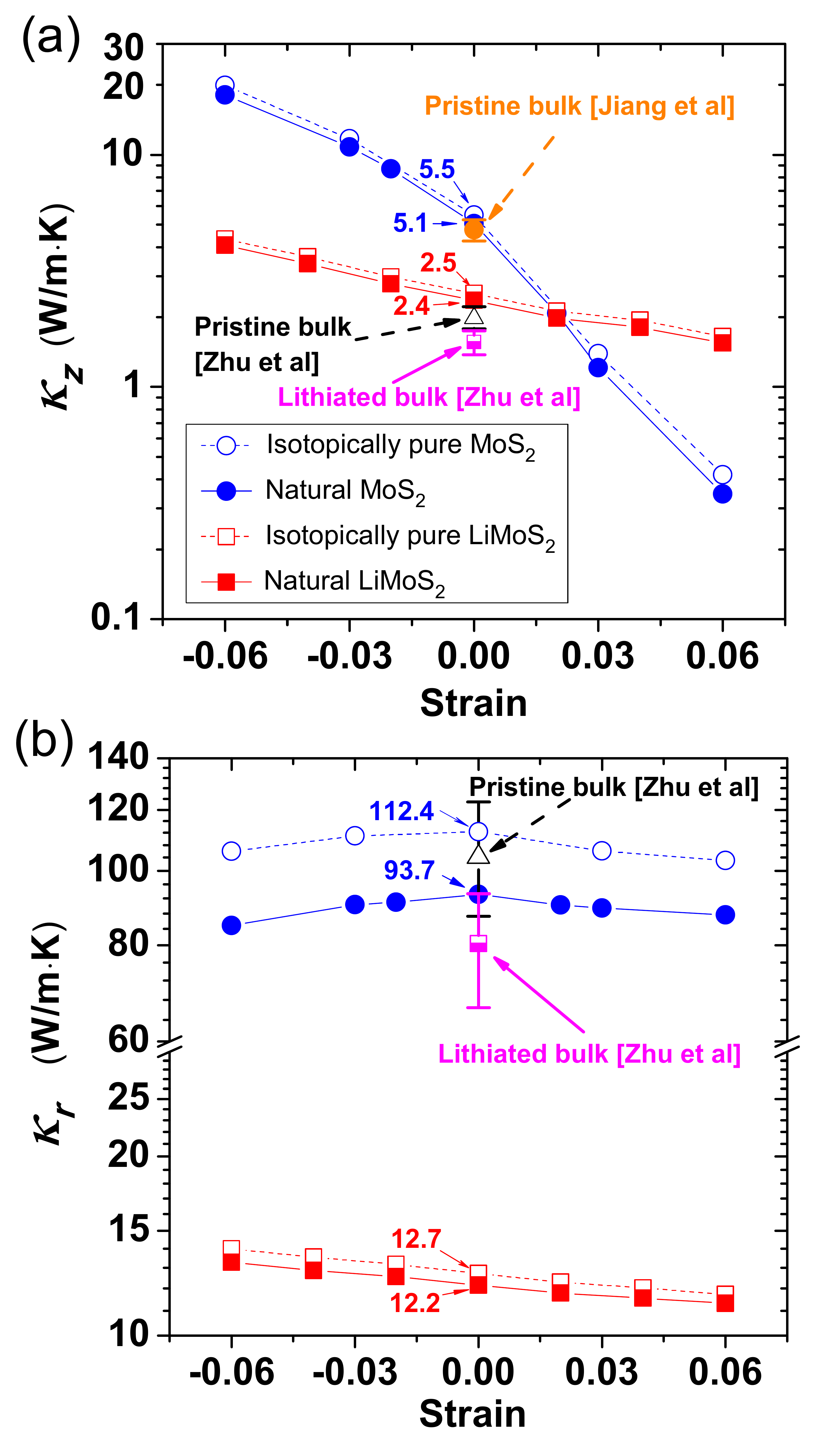}
\end{center}
	\caption{{Calculated} cross-plane (a) and in-plane (b) thermal conductivity as a function of cross-plane strain for pristine MoS\textsubscript{2} (blue open circles for isotopically pure \mos, blue full circles for natural \mos) and LiMoS\textsubscript{2} (red open squares for isotopically pure Li\mos, red full squares for natural Li\mos), also showing  measurements from Zhu \etal \cite{zhu_tuning_2016} and Jiang \etal \cite{jiang_time-domain_2017,jiang_probing_2017}.}
	\label{fig:kappa}
\end{figure*}


\paragraph{Structure.} The stable 2H phase of \Mos consists of planes with one formula unit each per unit cell, arranged in AB stacking.  
Upon  lithium intercalation, MoS\textsubscript{2} planes undergo a transition from trigonal prismatic 2H  to distorted octahedral
1T, and their stacking changes from AB to AA \cite{cheng_origin_2014}. The in-plane symmetry is reduced and in the 1T phase the primitive cell contains {four} Li\Mos units in a $2 \times 2$ superstructure. 

The optimized lattice parameters of both pristine and lithiated \Mos (Table \ref{table}) compare very well with experiments. 
These results confirm that our computational framework describes accurately the structural features of \Mos and their changes upon Li intercalation. 
In particular, we note that the 2H to 1T transition, occurring upon intercalation, produces an expansion of the in-plane lattice parameter {by} 7.8\% \cite{mulhern_lithium_1989}, but accommodates lithium with an inter-planar expansion as small as 0.5\%, as also observed in recent experiments \cite{zhu_tuning_2016}.
vdW-DF-cx calculations give lattice constants closer to experiments (within 0.4\%) than LDA, and consistent lattice expansion rates: a-axis {(in-plane)} expansion of 7.5\% and  c-axis {(cross-plane)} expansion of 1.4\%, which are in good agreement with experimental data \cite{mulhern_lithium_1989}.
This result is important, because it suggests that in experiments, in which large c-axis expansion rates are observed \cite{somoano_alkali_1973,py_structural_1983,rocquefelte_synergetic_2003}, either the transition to the 1T phase is not complete or samples entail mesoscale disorder that induces strain.

Upon intercalation Li donates an electron to the system, which would in principle become metallic. However a Jahn-Teller transition \cite{jahn_stability_1937} to the distorted 1T phase re-instates a gap between the valence and conduction band, making the system a semiconductor. Even if the 1T lithiated system {were} assumed to be metallic, the electrical contribution to thermal conductivity would be much smaller than the lattice contribution, as estimated by Zhu \etal \cite{zhu_tuning_2016}. Hence, in the following we can safely neglect the contribution of electrons to thermal transport. 

\paragraph{Thermal conductivity.} 
Figure \ref{fig:kappa} shows the in-plane and cross-plane thermal conductivity of pristine and fully lithiated \Mos as a function of strain.  
Since the discrepancies among former FP-BTE calculations \cite{gandi_thermal_2016,gu_layer_2016,lindroth_thermal_2016} of the thermal conductivity of pure \Mos may be due to convergence issues, we have taken care of converging our calculations with respect to all the DFT and BTE parameters (see Supplementary Figures S1-S13). In particular, due to the long-range nature of the interactions in \mos, it is critical to consider interactions up to the 9th neighbors shell to construct the anharmonic force constants tensor in order to achieve well-converged results for $\kappa$ (Figure S7). 

We find that  {isotopic scattering has a strong effect on $\kappa_r$, for which isotopically pure MoS\textsubscript{2} turns out about 20\% higher  than that of MoS\textsubscript{2} with natural composition. The isotopic effect for $\kappa_z$ is milder as the difference is only 10\%, and it is even weaker on both $\kappa_r$ and $\kappa_z$ of LiMoS\textsubscript{2} (see Figure 1 and Figure S25 for more details). The following results refer to materials with natural isotopic composition.}

{Along} with structural changes, lithium intercalation into MoS\textsubscript{2} leads to a seven-fold reduction of in-plane thermal conductivity, and two-fold reduction in cross-plane thermal conductivity. The two-fold reduction in $\kappa_{{z}}$ upon full lithiation predicted by FP-BTE is on the same order as the reduction of $\kappa_{{z}}$ measured by Zhu \etal \cite{zhu_tuning_2016} for bulk \Mos (the measured $\kappa_{{z}}$ drops from $\sim$2 to $\sim$1.6 \wmk\ upon lithiation to $x=0.86$, a factor of $\sim$1.25). 
Note that our FP-BTE predicts $\kappa_{{z}}$ of pristine bulk \Mos of about $\sim$5 \wmk,  whereas Zhu {\it et al} \cite{zhu_tuning_2016} measured $\sim$2 \wmk. Such discrepancy may be due to the small penetration depth (high modulation frequency) of the time-domain thermoreflectance (TDTR) measurement, as discussed in Refs.~\cite{jiang_time-domain_2017,jiang_probing_2017,ballisticMoS2}.
However, given that the source of \Mos crystals is generally mineralogical, variations in crystal quality between samples cannot entirely be ruled out.
Our results set the theory benchmark for the thermal conductivity of an unlithiated pristine system, and are in excellent agreement with recent measurements by Jiang \etal \cite{jiang_time-domain_2017,jiang_probing_2017} and DFT calculations by Lindroth \etal \cite{lindroth_thermal_2016}.
As for in-plane thermal conductivity there is a striking discrepancy between calculations and TDTR measurements \cite{zhu_tuning_2016}, which suggest that $\kappa_{{r}}$ is only slightly affected by lithium intercalation. 
Such discrepancy between {\sl ab initio} theory and experiments would deserve further investigation.


\begin{figure*}[t]
\begin{center}
\includegraphics[scale=0.31]{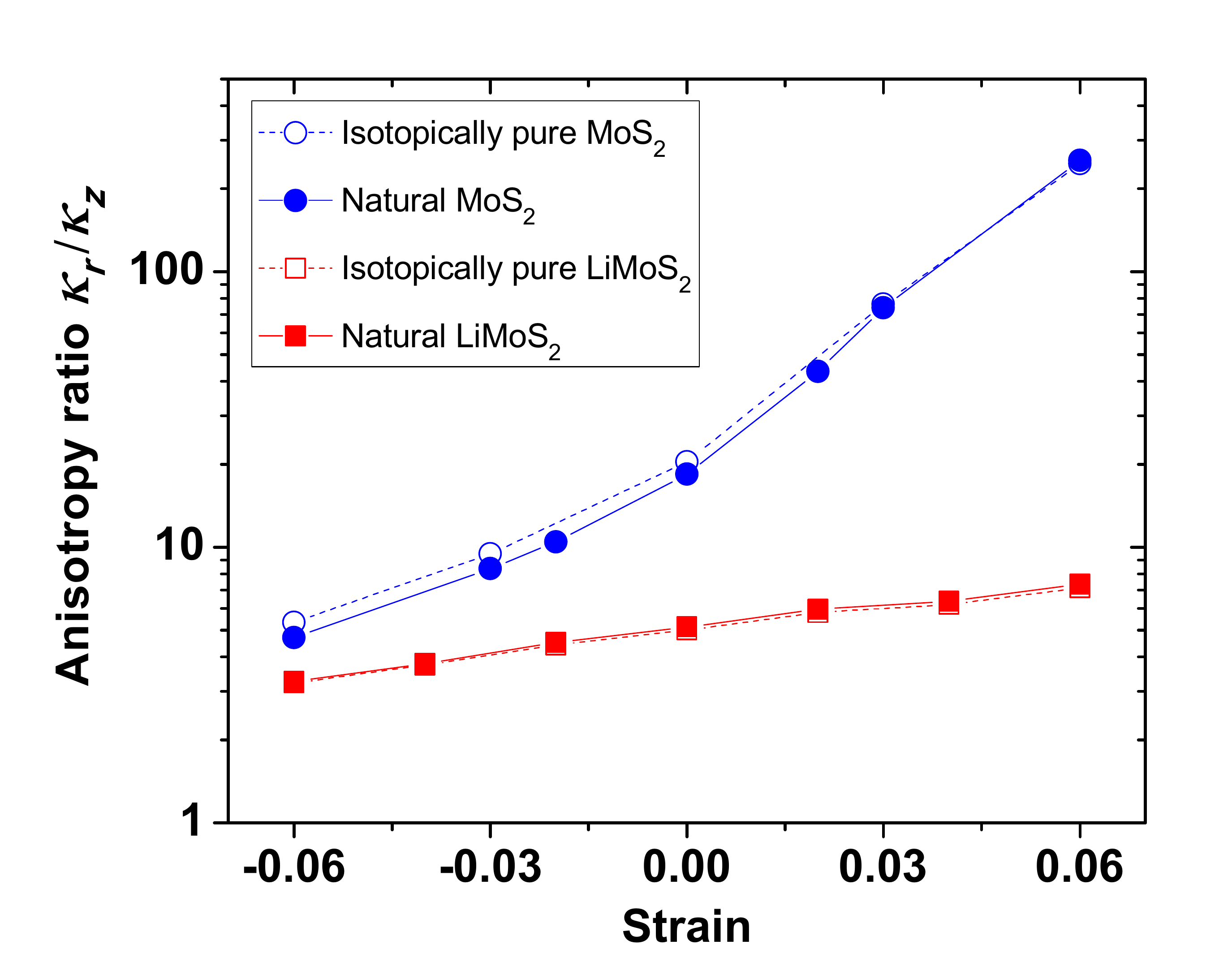}
\end{center}
\caption{Anisotropy ratio {of} in-plane thermal conductivity $\kappa_{{r}}$ {to} cross-plane thermal
conductivity  $\kappa_{{z}}$ as a function of cross-plane strain for pristine MoS\textsubscript{2} (blue open circles for isotopically pure \mos, blue full circles for natural \mos) and LiMoS\textsubscript{2} (red open squares for isotopically pure Li\mos, red full squares for natural Li\mos). The stress corresponding to the strain considered is reported in  Figure S14.}
\label{fig:anisotropy}
\end{figure*}


Even larger modulations of $\kappa$ occur in samples strained in the direction perpendicular to the \Mos planes.
In these calculations the cross plane lattice parameter $c$ is fixed and all the other parameters, including the in-plane lattice parameter $a$ are relaxed.
We consider strain ($\varepsilon$)  from -6\% (compressive) to $6\%$ (tensile), corresponding to applied stress from -6 to 5 GPa, respectively (Figure S14).
The cross-plane thermal conductivity of pristine MoS\textsubscript{2} varies monotonically over two orders of magnitude, from 0.35 \wmk\ upon $6\%$ tensile strain to 18.09 \wmk\ upon $6\%$ compressive strain. 
 {Recent measurements show that $\kappa_{{z}}$ increases under compressive strain in multilayer \mos\ \cite{exp_kappaz_strain_arxiv}, in qualitative agreement with our predictions.} 
In contrast, the cross-plane thermal conductivity of Li\Mos exhibits less variability, from 1.5 \wmk\ upon $6\%$ tensile strain to 4.1 \wmk\ upon $6\%$ compressive strain. Consequently, a crossover in $\kappa_{{z}}$ occurs upon tensile strain: the cross-plane thermal conductivity of Li\Mos becomes larger than that of \Mos for tensile strain larger than 2$\%$. 
Cross-plane strain also causes variations of the in-plane thermal conductivity but to a lesser extent. $\kappa_{{r}}$ of \Mos is reduced by up to about 10$\%$  by either tensile or compressive strain. $\kappa_{{r}}$ of Li\Mos varies monotonically from 11.3 \wmk\ for systems subject to large tensile strain to 13.3 \wmk\ upon compressive strain.

Since cross-plane and in-plane components of the thermal conductivity tensor show different sensitivity to cross-plane strain, it is then possible to modulate the anisotropy ratio ($\kappa_{{r}}/\kappa_{{z}}$) over a very wide range of values. Within the strain range investigated, the anisotropy ratio of \Mos is varied from $\sim$5
to $\sim$250 (Figure~\ref{fig:anisotropy}).  
Conversely, lithiation significantly reduces the overall anisotropy of $\kappa$ and the ratio $\kappa_{{r}}/\kappa_{{z}}$ changes from 3.3 to 7.3 as a function of strain. {The modulation of the anisotropy ratio of $\kappa$ is not significantly affected by isotopic composition.}


\begin{figure}[t]
\begin{center}
\includegraphics[scale=0.29]{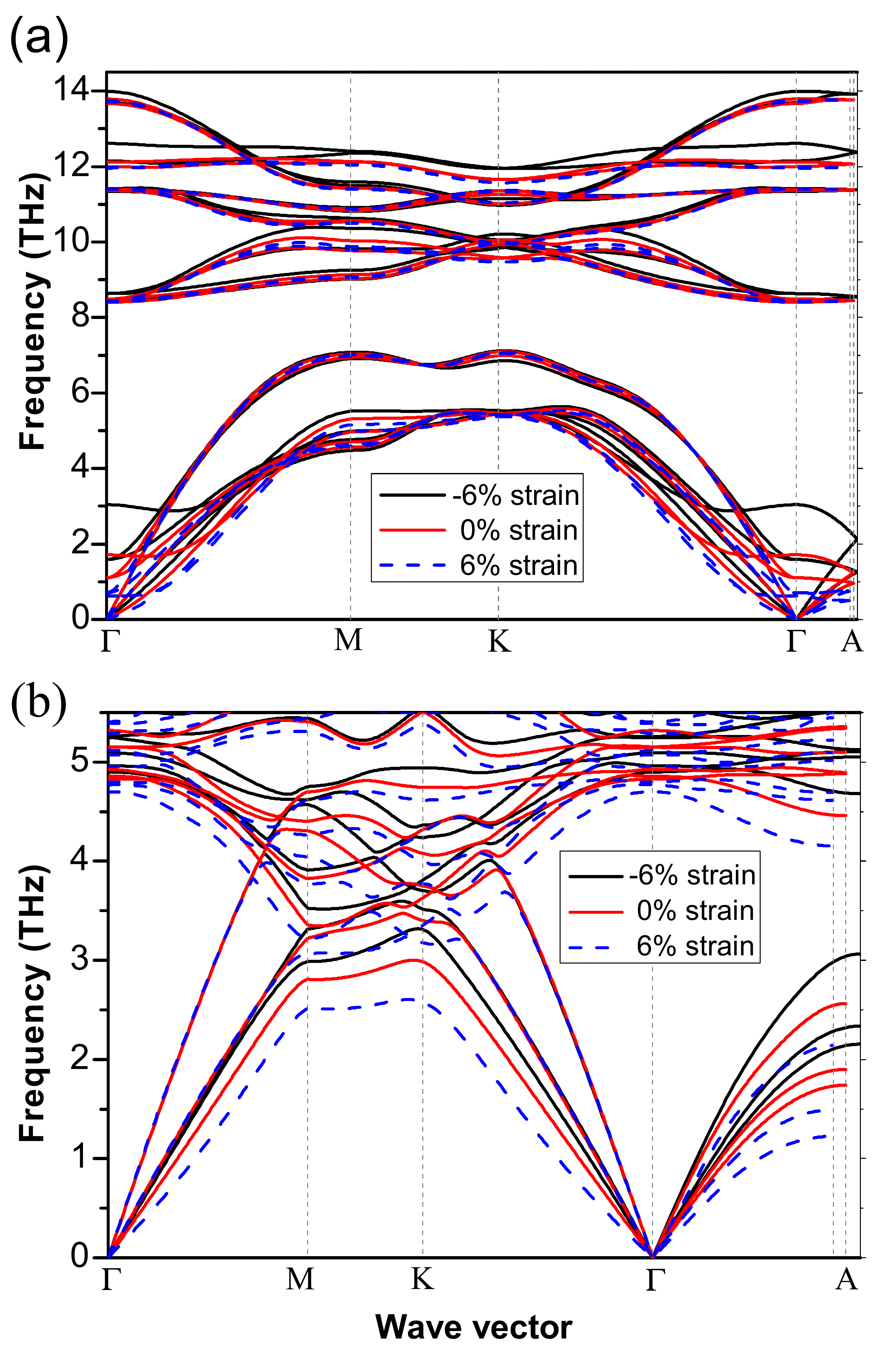}
\end{center}
\caption{Phonon dispersion relations for (a) MoS\textsubscript{2} and (b) LiMoS\textsubscript{2}, under -6\% (black lines), 0\% (red lines), and 6\% strain (blue dashed lines).} 
\label{fig:dispersion}
\end{figure}


\begin{figure*}[t]
\includegraphics[scale=0.62]{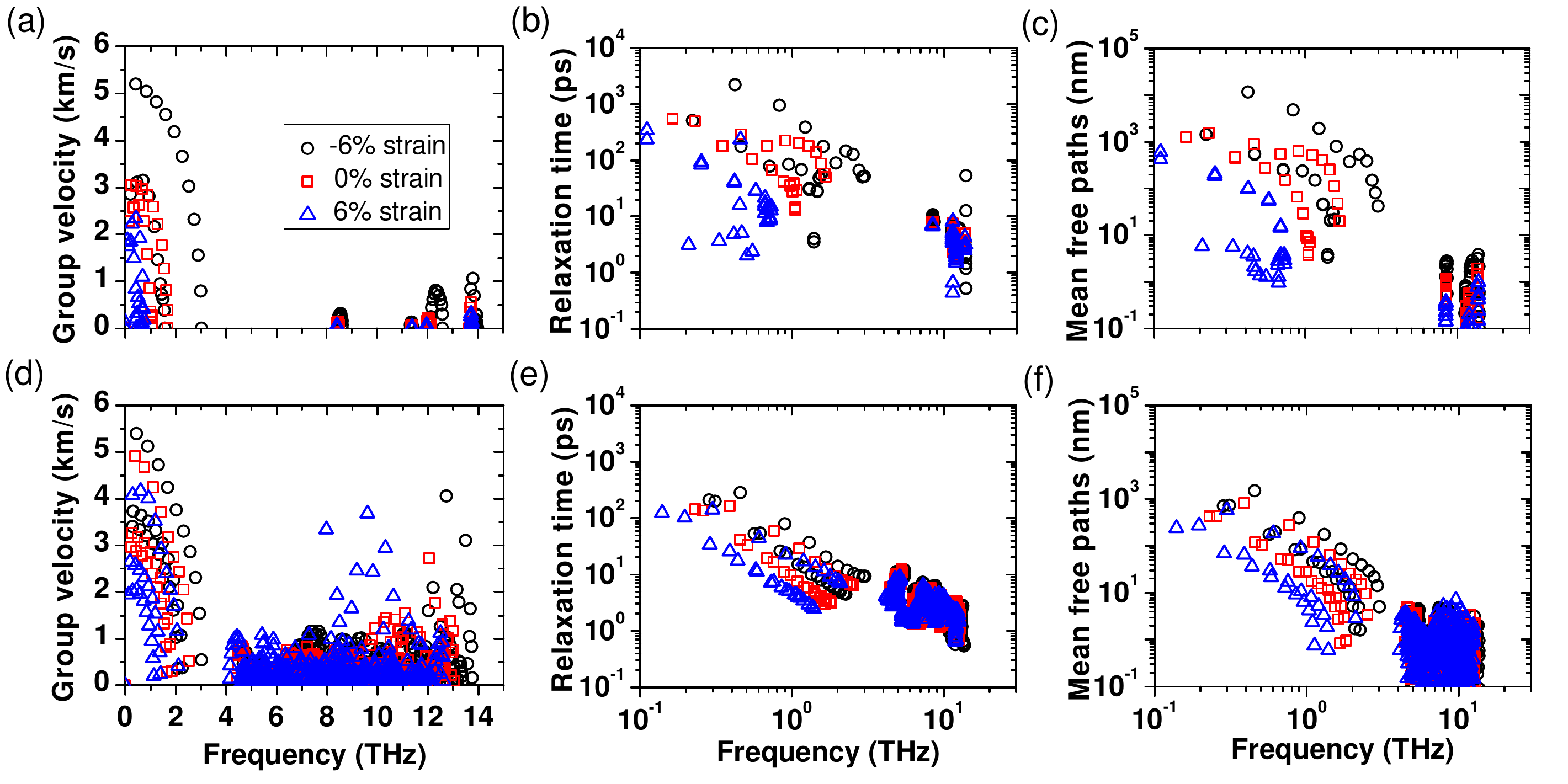}
\caption{(a-c) Phonon group velocity, relaxation time, and mean free paths along $\Gamma$-A
(cross-plane) direction for pristine MoS\textsubscript{2} under -6\%
(black circles), 0\% (red squares), and 6\% strain (blue triangles);
(d-f) the same but for LiMoS\textsubscript{2}.}
\label{fig:analysis-cross}
\end{figure*}

\begin{figure*}[t]
\includegraphics[scale=0.62]{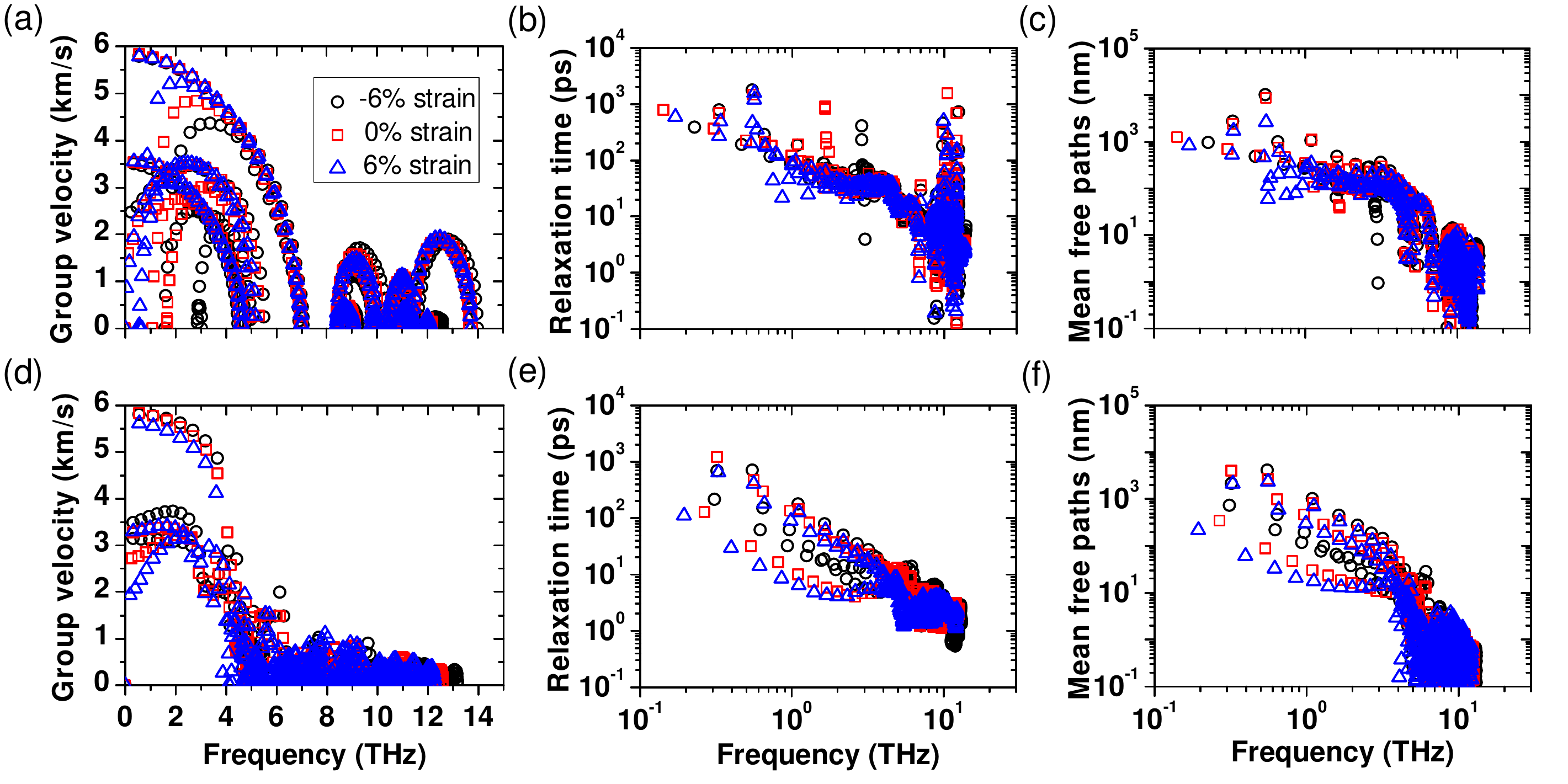}
\caption{(a-c) Phonon group velocity, relaxation time, and mean free paths along $\Gamma$-M
(in-plane) direction for pristine MoS\textsubscript{2} under -6\% (black circles),
0\% (red squares), and 6\% strain (blue triangles); (d-f) the same
but for LiMoS\textsubscript{2}.}
\label{fig:analysis-in}
\end{figure*}

\begin{figure*}
\includegraphics[scale=0.58]{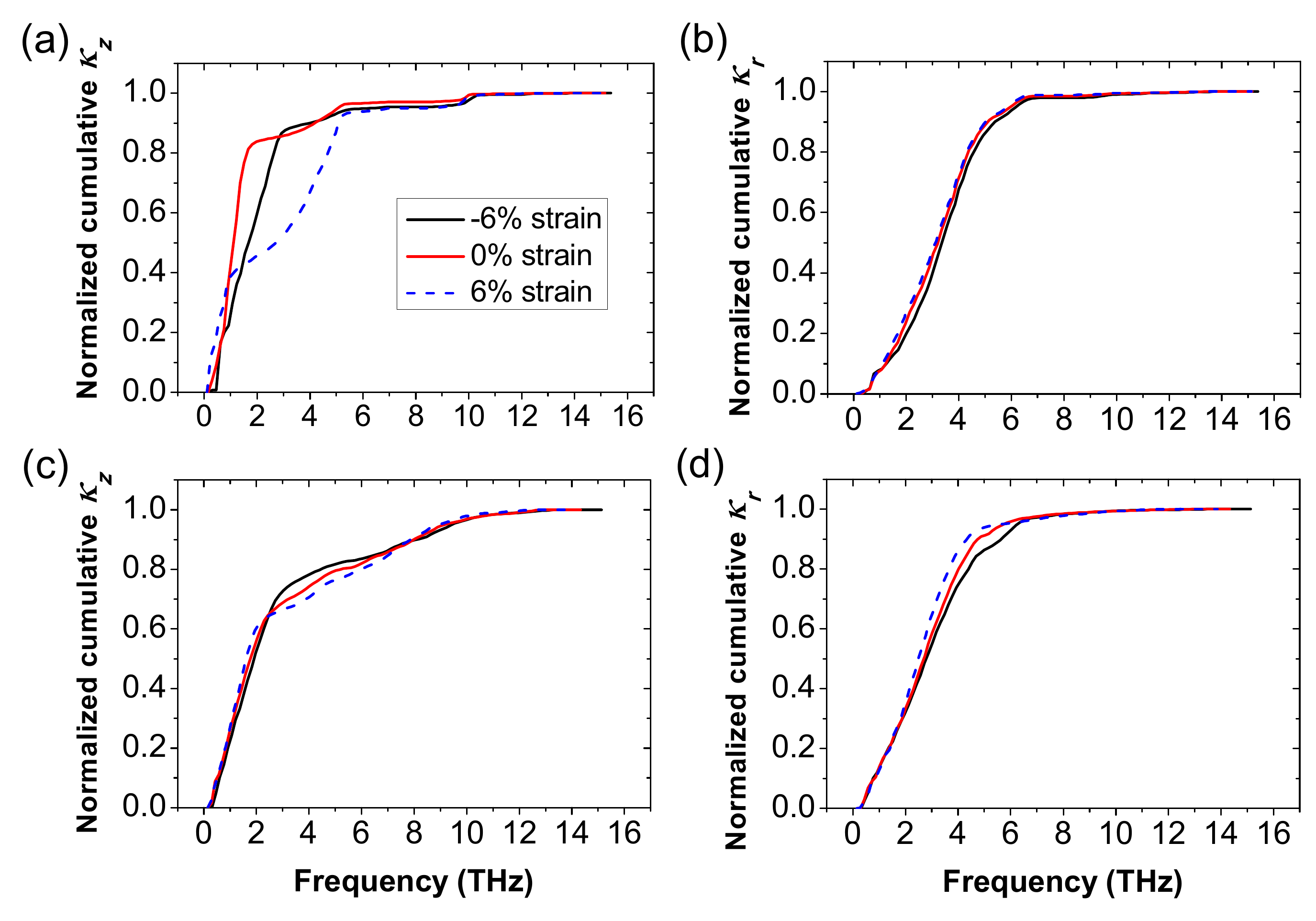}
\caption{Normalized frequency-dependent cumulative cross-plane (a) and and in-plane (b) thermal
conductivity for pristine MoS\textsubscript{2} under -6\% (black line), 0\% (red line), and 6\% strain (blue dashed line); (c-d) the same but for LiMoS\textsubscript{2}.}
\label{fig:freq-cumulative}
\end{figure*}

\paragraph{Phonon properties.}
We now analyze the origin of the modulation of thermal conductivity induced by lithium intercalation and strain, in terms of phonon properties and their contribution to $\kappa$, as expressed by Eq.~\ref{Eq.kappa}.
The two panels of Figure~\ref{fig:dispersion} display the effects of Li intercalation and strain on phonon dispersion relations. 
Previous reports showed that the majority heat carriers are acoustic modes with frequency below 6 THz for in-plane and 3 THz for cross-plane (see also Figure~\ref{fig:freq-cumulative}), and that the gap between acoustic and optical modes in pure \Mos limits the amount of viable scattering channels, leading to very high in-plane thermal conductivity \cite{gandi_thermal_2016,gu_layer_2016,lindroth_thermal_2016}. 

Upon intercalation we observe two major changes in the dispersion relations. On the one hand, both in-plane and cross-plane acoustic modes are  stiffened upon lithiation. On the other hand the larger number of atoms per unit cell in Li\Mos engenders a corresponding number of phonon branches that fill the gaps of the spectrum of MoS\textsubscript{2} (see Figure~\ref{fig:dispersion}, and Figures S12, S13 and S15).
The majority of these optical modes, especially those localized on Li atoms have relatively flat dispersion and do not carry significant amounts of heat, but they contribute scattering channels to acoustic modes \cite{ECT-Transistor}. The overall effect is a significant reduction of $\kappa$, especially in plane.

Cross-plane strain mostly influences the dispersion relations of the acoustic modes of both \Mos and LiMoS\textsubscript{2}.
Along the $\Gamma$-A high symmetry direction, i.e. cross-plane, compressive strain induces a significant stiffening of all three acoustic branches in both systems, while tensile strain induces softening.  
For phonons propagating in-plane ($\Gamma$-M-K-$\Gamma$ path), cross-plane strain impacts only the flexural modes, producing either stiffening (compressive) or softening (tensile). 

The magnitude of such changes is much larger for \Mos than for LiMoS\textsubscript{2}, as confirmed by the calculation of the cross plane group velocities, reported in Figure \ref{fig:analysis-cross}a and \ref{fig:analysis-cross}d. The group velocities of the phonons propagating across the layers of  \Mos are largely enhanced upon compression and reduced upon tensile strain, in accordance with the effect seen on its thermal conductivity. In particular, the speed of sound increases from 3000 m/s to 5500 m/s upon 6$\%$ compression. 
Considering that acoustic modes are the main heat carriers, and that $\kappa$ is proportional to $v^2$, such change is sufficient to justify the increase of thermal conductivity upon compression from $\sim 5$ to $\sim 18$ \wmk\ (Figure~\ref{fig:kappa}a). 
The same argument explains the reduction of $\kappa_{{z}}$ upon tensile strain.

Such variations of group velocities are accompanied by enhancement/reduction in phonon relaxation times (Figure~\ref{fig:analysis-cross}b), hence in mean free paths ({Figure}~\ref{fig:analysis-cross}c), which overall contribute to the observed strong modulation of $\kappa$. 
In LiMoS\textsubscript{2}, group velocities, relaxation times and mean free paths follow the same trends as in MoS\textsubscript{2}, but the effect of strain is much weaker (Figure~\ref{fig:analysis-cross}d-f). This is because the stiffening of the structure upon intercalation limits the range of variation of the cross plane group velocities. For compressed and unstrained systems, phonon lifetimes in LiMoS\textsubscript{2} are significantly smaller than those in pristine \mos, resulting in a much reduced $\kappa_{{z}}$. However, the smaller range of variation of group velocities and lifetimes upon tensile strain leads to the observed crossover in $\kappa_{{z}}$ at $2\%$ strain. 
Hence intercalated metals may  play a role as channels between layers, possibly enhancing heat transport in weakly bonded strained systems.

The analysis of the in-plane phonon properties of \Mos and Li\Mos sheds light on the trends of $\kappa_{{r}}$. Group velocities, relaxation times and mean free paths along the high symmetry $\Gamma$-M direction are reported in Figure {\ref{fig:analysis-in}a-c} for pristine MoS\textsubscript{2} and {Figure \ref{fig:analysis-in}d-f} for LiMoS\textsubscript{2}.
The effect of lithium intercalation on in-plane phonon modes is to reduce the frequency range of acoustic modes with significant group velocity, as well as to reduce the overall lifetimes and mean free paths. Both these effects are consequences of the 2H--1T structural transition, and account for the observed major reduction of $\kappa_{{r}}$ from 93.2 to 12.2 \wmk. 
For pristine MoS\textsubscript{2}, strain reduces the MFPs of the ZA modes through two different mechanisms. Compressive strain reduces the group velocities, whereas tensile strain reduces the relaxation times. These two disparate effects cause a mild reduction of $\kappa_{{r}}$ upon both tensile and compressive cross-plane strain, inducing the non-monotonic behavior shown in Figure~\ref{fig:kappa}b.  
For LiMoS\textsubscript{2}, both the group velocities and the relaxation times of the ZA modes increase upon compressive deformations and decrease under tensile strain, leading to a monotonic trend of $\kappa_{{r}}$ versus strain.

These effects modify the relative contribution of phonons with different frequency to the total cross-plane and in-plane thermal conductivity, as shown by the cumulative $\kappa_{{r}}$ and $\kappa_{{z}}$ as a function of frequency reported in Figure~\ref{fig:freq-cumulative}.
For pristine \mos, compressive strain extends the contribution of higher frequencies to $\kappa_{{z}}$ by stiffening of the acoustic branches.
Conversely, tensile strain reduces the range of the acoustic branches so much that the relative contribution of higher frequencies appears more relevant. Li intercalation extends substantially the range of frequencies that contribute to cross-plane heat transport, making the contributions from optical modes up to $\sim 11$ THz relevant. As for in-plane heat transport, Li intercalation significantly affects the absolute value of $\kappa_{{r}}$, but not the relative contribution of different phonon frequencies. 
The relative contribution of different phonon frequencies to $\kappa_{{r}}$ is mildly affected for the lightly stiffening (softening) of flexural modes by compressive (tensile) strain.

 {
Although we have not considered the effect of in-plane strain, we argue that it would induce the modulation of phonon transport analogous to those computed upon cross-plane strain, due to the mechanical response of the system through its Poisson's ratio (Figures S26, S27) \cite{Poissons_NatMat_2011,Poisson_NatCom_2014,Poissons_PRB_2016}.
Consistently with this argument, MD simulations suggested that both tensile and compressive in-plane strain would reduce $\kappa_r$ of single-layer \mos\ \cite{ding_manipulating_2015}.
}

\section{Conclusions}

In conclusion, {first-principles calculations performed with strict convergence criteria allow} us to resolve discrepancies in the literature about the thermal conductivity of \mos, thus establishing a reliable benchmark for future experiments and simulations.

These calculations show that both the in-plane and cross-plane thermal conductivity of \Mos are critically sensitive to both strain and intercalation. 
We predict that lithium intercalation of bulk \Mos produces more than seven-fold reduction in $\kappa_{{r}}$, and two-fold reduction in $\kappa_{{z}}$. These results constitute a lower bound for the thermal conductivity modulation caused by Li intercalation, without considering structural disorder (mixed phases, vacancies, etc).
Most remarkably, $\kappa_{{z}}$ of MoS\textsubscript{2} can be modulated over two orders of magnitude by strain {between} -6 to 6$\%$, whereas $\kappa_{{z}}$ of LiMoS\textsubscript{2} is less sensitive to strain. By combining strain and intercalation it is possible to tune the anisotropy ratio of $\kappa$ from 5 to 250. 
Being able to modulate the thermal conductivity anisotropy ratio over such a wide range can have important applications in thermal management.
These calculations provide a theoretical benchmark that will enable the interpretation of experimental measurements of thermal conduction in intercalated van der Waals transition metal dichalcogenides. They provide guidelines to design materials and engineer devices with tunable thermal conductivity.




\section{Methods}

 To properly take into account collective effects in phonon transport, we solve the BTE self-consistently, as implemented in the ShengBTE code \cite{sparavigna_lattice_2002,li_thermal_2012,ward_ab_2009,li_shengbte:_2014}.

DFT calculations are performed within local density approximation (LDA) of the exchange and correlation functional \cite{perdew_self-interaction_1981} by using the Quantum-Espresso package \cite{giannozzi_quantum_2009,QE-2017}. 
Core electrons are approximated using norm-conserving pseudopotentials \cite{hartwigsen_relativistic_1998}, and the valence electronic wavefunctions are expanded in a plane-wave basis set with a kinetic energy cutoff of 100 Ry.
The charge density is integrated on 10x10x4 and 4x4x4 Monkhorst-Pack meshes of k-points for pristine MoS\textsubscript{2} and LiMoS\textsubscript{2}, respectively. Structural and cell relaxations are performed using 
a quasi-Newton optimization algorithm with a tight convergence criterion of $10^{-8}$ Rydberg/Bohr for maximum residual force component. 
To model lithium intercalation of MoS\textsubscript{2}, we used the optimized \Mos structure to construct a 2x2x1 supercell, and inserted Li atoms in the vdW gap. We verified that the system relaxes spontaneously to the distorted 1T phase. 

We utilize density-functional perturbation theory (DFPT) \cite{baroni_phonons_2001} to calculate harmonic interatomic force constants (IFCs) with 10x10x4 and 4x4x4 q-point meshes for pristine (2H)-MoS\textsubscript{2} and LiMoS\textsubscript{2}, respectively.
Anharmonic third order force constants for the calculation of lattice thermal conductivity are computed by finite differences up to a cutoff interatomic distance 7.04 \AA, corresponding to the 11th nearest neighbour shell, in a $5\times5\times1$ supercell containing  $150$ atoms for pristine (2H)-MoS\textsubscript{2}. For (1T)-LiMoS\textsubscript{2} we use a cutoff up to 8th nearest neighbour in a $2\times2\times2$ super-cell with $128$ atoms.
Third derivatives of the potential energy with respect to displacements of atoms $ijk$ along $\alpha\beta\gamma$ coordinates are computed by finite differences  ($\delta x=0.01$ \AA), and translational invariance is enforced using the Lagrangian approach \cite{li_shengbte:_2014}. 


For pristine MoS\textsubscript{2}, it is necessary to take iterative self-consistent (SCF) calculations, since relaxation time approximation (RTA) underestimates its thermal conductivity (see Figures S16 and S17). For LiMoS\textsubscript{2}, we found that RTA yields the same results as SCF within 1\% (see Figures S18 and S19). We carefully checked the convergence with q-points grids up to 45x45x11 for pristine MoS\textsubscript{2} (see Figure S5) and up to 23x23x23 for LiMoS\textsubscript{2} (see Figure S6). Details on convergence tests are provided in the supporting information.

The main results of this work are confirmed by calculations performed using a van der Waals density functional (vdW-DF) \cite{thonhauser_spin_2015,thonhauser_van_2007,berland_van_2015,langreth_density_2009} with non-empirical consistent exchange (vdW-DF-cx) \cite{berland_exchange_2014,berland_van_2014}, and projector augmented wave (PAW) pseudopotentials \cite{blochl_projector_1994,kresse_ultrasoft_1999} (see Figures S20-S24). 
These tests justify {\it a posteriori} the use of LDA, which reproduces the van der Waals interaction among different planes due to error cancellation between the exchange and the correlation parts of the functional. More details are provided in the supporting information.


\noindent{\it Supporting Information\/}: Detailed convergence tests of FP-BTE calculations, full phonon dispersion curves, stress-strain curves for pristine MoS\textsubscript{2} and LiMoS\textsubscript{2} (Figures S1-S15); comparison between the self-consistent (SCF) solution and relaxation time approximation (RTA) of the Boltzmann transport equation (Figures S16-S19); calculations performed using van der Waals density functional (vdW-DF, with non-empirical consistent exchange) and projector augmented wave (PAW) pseudopotentials (Figures S20-S24);  {a comparison of $\kappa$ of systems with pure and natural isotopic composition upon cross-plane strain (Figure S25); the Poisson's ratio for pristine \mos\ and Li\mos\ by LDA and vdW-DF with non-empirical consistent exchange (Figures S26-S27).}





\ack
Useful discussions with Jes\'{u}s Carrete, Wu Li, Majid Zeraati, David Strubbe, Daniele Selli and Shruba Ganghopadhyay are gratefully acknowledged. AS, EP and KEG acknowledge funding from the National Science Foundation (NSF) under the EFRI 2-DARE grant 1542883.


\section*{References}
\bibliographystyle{iopart-num.bst}
\bibliography{bibtex.bib}
\end{document}